\documentclass[12pt,preprint]{aastex}

\slugcomment{To be published in ApJ }

\shorttitle{Time Domain Coherence}
\shortauthors{Che, Liu \&  Li}

\begin{document}

\title{X-ray Variability Coherence in the Time Domain}

\author{Xiao Che\altaffilmark{1}, Cong-Zhan Liu\altaffilmark{1} and Ti-Pei Li\altaffilmark{1,2,3}}

\altaffiltext{1}{Department of Engineering Physics \& Center for
Astrophysics, Tsinghua University, Beijing, China}
\altaffiltext{2}{Department of Physics \& Center for Astrophysics,
Tsinghua University, Beijing, China} \altaffiltext{3}{Particle Astrophysics Lab.,
Institute of High Energy Physics, Chinese Academy of Sciences, Beijing}

\begin{abstract}
A technique for calculation of variability coherence at different
timescales performed directly in the time domain is introduced.
Simulations are made to compare the coherence spectrum derived by the
time domain technique with the
coherence function by Fourier analysis. The results indicate
that in comparison with the Fourier analysis the time domain
technique is more powerful for revealing signal coherence in noisy data.
We apply the time domain technique to the real data of the black
hole binaries Cygnus X-1 and GX 339-4 and compare the results with
their Fourier coherence spectra.
\end{abstract}


\section{Introduction}
 The complex aperiodic variability of high-energy emission
shown in different time scales is a common character for X-ray binaries
and super massive black
 holes. Temporal analysis is an important approach to study undergoing
physics in objects from observed light curves. The spectral lag
between variation signals in different energy bands provides useful
information on their producing and propagating processes in
celestial objects. Besides the time lag, the spectral coherence, a
measure of the degree of linear correlation between two light curves
in different energy bands, is another useful quantity to provide
constraints on models of physical processes in observed objects. For
studying variability coherence in the frequency domain, the
coherency function -- the normalized average amplitude of the cross
power density as a function of Fourier frequency -- can be used.
Vaughan and Nowak (1997) have presented a procedure for estimating
the coherence function in the presence of counting noise, which has
been applied to studying black hole candidates (e.g. Nowak \&
Vaughan 1996; Vaughan \& Nowak 1997; Cui, Zhang \& Chen 2000; Nowak
et al. 2001; Nowak, Wilms \& Dove 2002; Ji et al. 2003).

The time domain quantity corresponding to the coherence in the frequency domain
is the coefficient of linear correlation between two light curves
in different energy bands.
The correlation coefficient has not been extensively applied in
temporal analysis as the coherence function in the frequency domain.
One of the reasons is that the variability is caused by various physical processes
at different timescales,
a single value of correlation coefficient
from two observed light curves is not enough to provide measures of
association between complex variables. What needed is a distribution
of correlation coefficient over timescales. Here we present a
procedure for deriving the coherence spectrum in the time domain, or
a timescale spectrum of the correlation coefficient for a complex process
in the presence
of background noise and compare the coherence spectrum in the time
domain with the Fourier frequency-dependent function of coherence by
simulation in \S 2. In our simulation, the result from the time
domain technique is consistent with the coherence function obtained
from the Fourier analysis when the signal to noise ratio is high. In
general, the time domain technique can provide a more accurate and
stable result than the Fourier analysis. We apply the time domain
technique to the black hole X-ray binaries Cygnus X-1 and GX 339-1
and present the results in \S 3 and give discussion on timelag effect
in \S 4.

\section{Timescale Spectrum of Correlation Coefficient}
A proper statistic for measuring the degree of linear correlation
between variations in two $N$-point signal counting series $x_s(i)$
and $y_s(i)$ is their correlation coefficient
\begin{equation}
r=\frac{\sum_{i=1}^N(x_s(i)-\overline{x}_s)(y_s(i)-\overline{y}_s)/N}{\sigma(x_s)\sigma(y_s)}.
\end{equation}
For two observed lightcurves $x(i)=x_s(i)+x_n(i)$ and $y(i)=y_s(i)+y_n(i)$
with independent noise $x_n$ and $y_n$ respectively, we have
\[ \sum_i(x(i)-\overline{x})(y(i)-\overline{y})=
\sum_i(x_s(i)-\overline{x}_s)(y_s(i)-\overline{y}_s), \]
and \[
\sigma^2(x)=\sigma^2(x_s)+\sigma^2(x_n), \hspace{4mm} \sigma^2(y)=\sigma^2(y_s)+\sigma^2(y_n),
\]
then the signal's correlation coefficient
can be calculated as
\begin{equation}
  r=\frac{\sum_{i=1}^N(x_s(i)-\overline{x}_s)(y_s(i)-\overline{y}_s)/N}{\sigma(x_s)\sigma(y_s)}
=\frac{\sum_{i=1}^Ni(x(i)-\overline{x})(y(i)-\overline{y})/N}{\sqrt{\sigma^2(x)-\sigma^2(x_n)}
{\sqrt{\sigma^2(y)-\sigma^2(y_n)}}}~.
\end{equation}
It should be pointed out that one can not take two so called
``background-subtracted lightcurves'' as signal's lightcurves $x_s$ and $y_s$
to calculate the correlation coefficient simply by Eq.~(1), because background
lightcurves given by a database are usually smoothed curves not including
the fluctuation of real noise at shorter timescales.

We apply a timescale analysis technique (Li, Feng \& Chen 1999; Li 2001; Li et al. 2004)
to calculate correlation coefficients for different timescales from two observed
lightcurves.
From an original lightcurve $x(j; \delta t) ~(j=1,2,\cdots)$ with a time resolution $\delta t$,
we can construct $M$ different lightcurves with the same larger time step $\Delta t = M \delta t$
by combining $M$ successive bins
\begin{equation}
x_{m}(i; \Delta t)=\sum_{j=(i-1)M+m}^{iM+m-1}x(j; \delta t)~, \hspace{5mm}
(m=1,\cdots,M)~,
\end{equation}
where the combination starts from the $m$th bin of the original lightcurve.
From two simultaneously observed counting series
$x(j;\delta t)$ and $y(j;\delta t)$, the correlation coefficient
$r(\Delta t)$ for a certain timescale $\Delta t=M_{\Delta t}\delta t$
can be calculated by
\begin{equation}
r(\Delta t)=\frac{1}{M_{\Delta t}}\sum_{m=1}^{M_{\Delta t}}
\frac{\sum_{i=1}^N(x_m(i)-\overline{x}_m)(y_m(i)-\overline{y}_m)/N}{\sqrt{\sigma^2(x_m)-\sigma^2(x_n)}
{\sqrt{\sigma^2(y_m)-\sigma^2(y_n)}}}~.
\end{equation}
The total observation period is divided into $L$ segments. The
correlation coefficient $r_k(\Delta t)$ is calculated by Eq.~(4) for
a pair of lightcurves on each time segment $k$ separately.
 The average $\overline{r}(\Delta t)=\sum_{k=1}^Lr_k(\Delta t)/L$
and its standard deviation $\sigma(\overline{r})$ are defined as the coherence
coefficient for the timescale $\Delta t$ and its error.

To check the procedure described above and compare it with the
Fourier coherence function we produce two 1000 s signal time series,
$x_s$ and $y_s$, with time bin $\delta t=1$~ms sampled from a
two-dimensional normal distribution
\begin{eqnarray}
p(x_s,y_s)&=&\frac{1}{2\pi\sigma(x_s)\sigma(y_s)\sqrt{1-\rho^2}}
\exp\left\{-\frac{1}{2(1-\rho^2)}
\left[\left(\frac{x_s-\langle x_s\rangle }{\sigma(x_s)}\right)^2\right.\right. \nonumber \\
& &\left.\left.-2\rho\frac{x_s-\langle x_s\rangle }{\sigma(x_s)}\frac{y_s-\langle y_s\rangle }{\sigma(y_s)}
+\left(\frac{y_s-\langle y_s\rangle }{\sigma(y_s)}\right)^2\right]\right\}~,
\end{eqnarray}
where $\langle x_s \rangle , \sigma(x_s), \langle y_s \rangle ,
\sigma(y_s)$ are means and standard deviations of $x_s$ and $y_s$
respectively, and $\rho$ is the correlation coefficient between
$x_s$ and $y_s$. By giving these parameters in Eq.~(5), we are able
to produce two signal time series with a certain correlation
coefficient. To simplify, we specify $\langle x_s \rangle=\langle
y_s \rangle=49$, $\sigma(x_s)=\sigma(y_s)=7$. The correlation
coefficient $\rho$ in Eq.~(5) is the theoretical value of coherence,
and it is used to compare with results of time domain coherence and
Fourier coherence function. We add independent Poisson noise $x_n$
and $y_n$ on the signal series $x_s$ and $y_s$ respectively to get
simulated lightcurves $x=x_s+x_n$ and $y=y_s+y_n$. Different signal
to noise ratio (SNR) could be set by adjusting the mean of
independent Poisson noise.
The simulated lightcurve is divided into ten segments, each  has a
length of 100 s. For each segment, coherences at different timescales
 are calculated by Eq.~(4) and Fourier coherence coefficients at different
frequencies are also computed with the procedure
described by Vaughan and Nowak (1997). Finally, the average
and its standard deviation are computed from the ten time domain coherences
and Fourier coherence coefficients separately.

 Figure~1 shows coherence
spectra from simulated light curves with $\rho=0.8$ and SNR = 5, 2
and 1 separately, where Fourier coherence function in the frequency
domain are transformed into the time domain by $time
scale=1/frequency$. From Fig.~1 we can see that, in comparing with
Fourier coherence functions, time domain coherence distributions are
more consistent and accurate. For the case of low SNR, Fourier
coherence functions fluctuate seriously and have a large error bar,
while the time domain procedure has a better capability to reveal a
given coherence, especially at short timescales. Figure 2 show the
results for a certain SNR = 1 and different $\rho=0.9$, 0.8 and 0.6
respectively, indicating that the time domain procedure is more
powerful to reveal lower measured coherence.

\begin{figure}[b]
\plotone{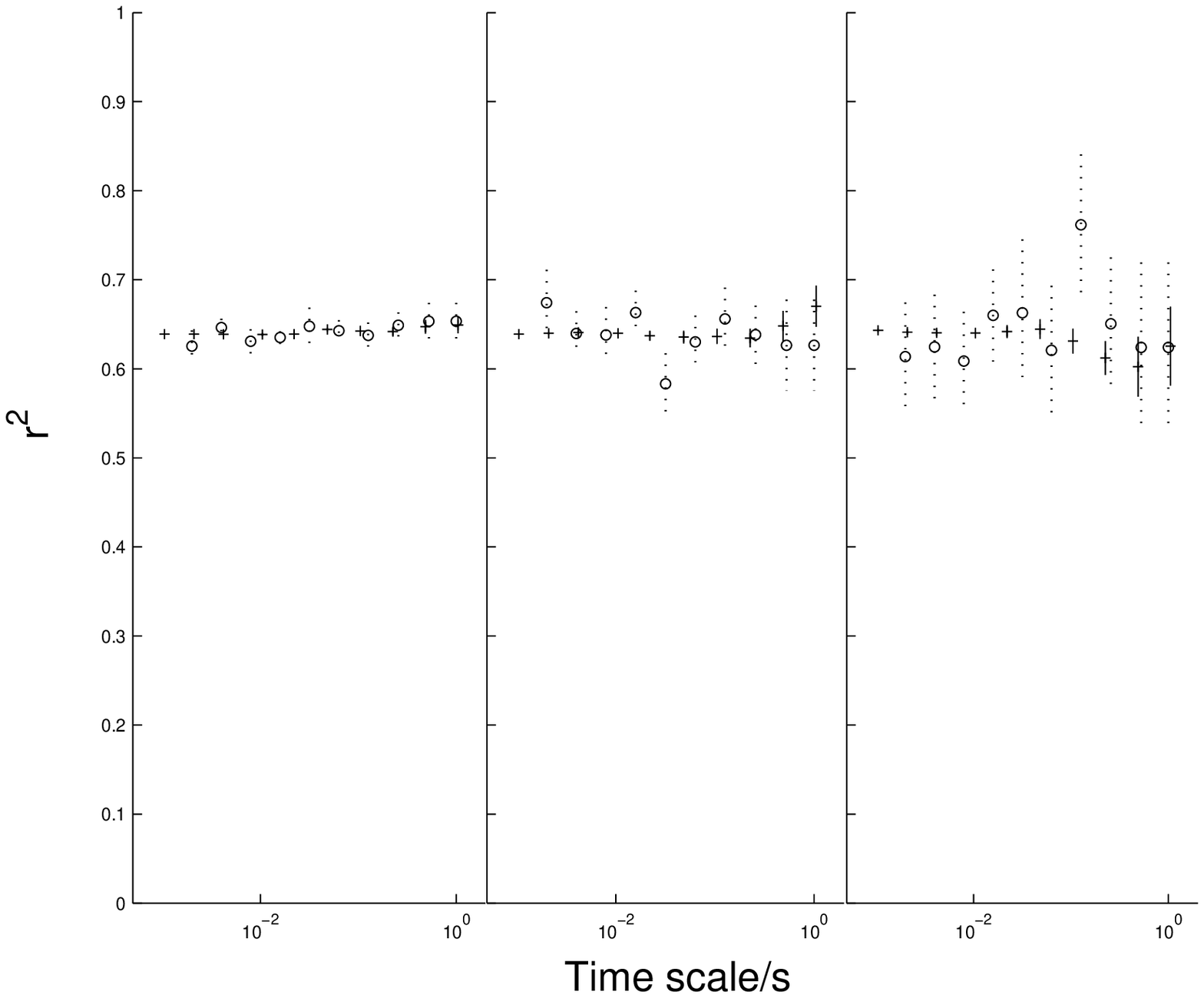} \caption{Coherence between two simulated
lightcurves with a given signal coherence $r^2=0.64$ and different SNR.
From left to right, SNR = 5, 2 and 1 respectively.
{\sl Plus sign} and {\sl solid line} represent the coherence and
its deviation obtained by the time domain technique.
{\sl Circle} and {\sl dotted line} show Fourier coherence function. \label{fig1}}
\end{figure}

\begin{figure}[b]
\plotone{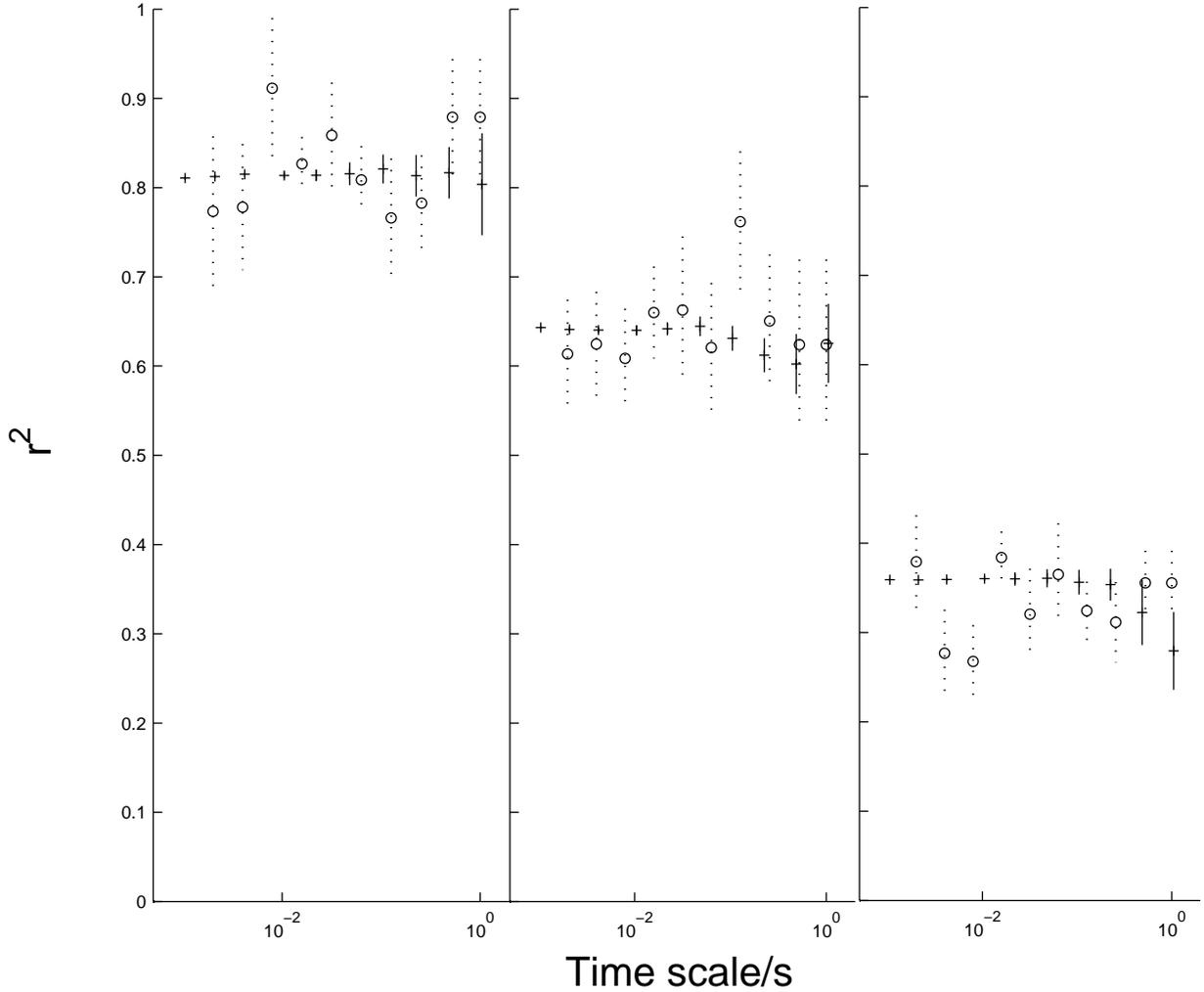} \caption{Coherence between two simulated
lightcurves with a given SNR = 1 and different signal correlation
coefficients. From left to right, the signal correlation coefficient
$r^2=0.81$, 0.64, and 0.36 respectively. {\sl Plus sign} and
{\sl solid line} represent the coherence and its deviation obtained
by the time domain technique. {\sl Circle} and {\sl dotted line}
show Fourier coherence function. \label{fig2}}
\end{figure}

The test and comparison are extended to different signal distributions
other than the binormal one.
The joint probability density function of binormal distribution can be factorized into two factors
\begin{equation}
p(x,y)=p(x)p(y|x)~,
\end{equation}
where
\begin{equation} p(x)=\frac{1}{\sqrt{2\pi}\sigma_{x}}\exp\left[-\frac{1}{2}\left(\frac{x-\langle
x\rangle}{\sigma_{x}}\right)^2\right]~,
\end{equation}
and
\begin{equation}
p(y|x)=\frac{1}{\sqrt{2\pi}\sigma_{y}\sqrt{1-\rho^2}}\exp\left[-\frac{1}{2}\left(\frac{y-\langle
y\rangle-\rho\frac{x-\langle x\rangle }{\sigma_{x}}\sigma_{y}
}{\sigma_{y}\sqrt{1-\rho^2}}\right)^2\right]~.
\end{equation}
 The first factor $p(x)$ of Eq.~(6) is a one-dimensional Gaussian density function of the variable
$x$ which is independent of $y$. The second factor $p(y|x)$ is also
a Gaussian density but with average $\overline
y+\rho\frac{x-\overline x }{\sigma_{x}}\sigma_{y}$, standard error
$\sigma_{y}\sqrt{1-\rho^2}$ and the correlation coefficient $\rho$
between x and y, which can be seen as a conditional density of $y$
given $x$. With Eq.~(6) we can produce a pair of time series with a given
correlation coefficient $\rho$ with a density function
$p(x)$ different to Gaussian by sampling $x$ from the density
$p(x)$ and $y$ from the density of Eq.~(8). The exponential
distribution, uniform distribution and Poisson distribution are
adopted as $p(x)$ to produce correlated pairs of time series
separately, and then compute their coherence distributions in the
time domain and their Fourier correlation functions. All results are
similar with Fig.~1 and Fig.~2, showing that the time domain
procedure is an effective approach to coherence analysis.

\section{X-ray Coherence of Black Hole Binaries}
Here we apply the time domain technique of coherence analysis to two
black hole candidates, Cyg X-1 in the low state and GX 339-4 in the
low and high states. We use observational data from PCA/$RXTE$.
PCA/$RXTE$ observation has dead time $10\mu s$ much smaller than
time resolution 1ms of original lightcurves we use in this paper.
And for Cyg X-1, dead-time affection is about $1\%$ (Maccarone and
Coppi 2000), so it could be ignored.

Variabilities in lightcurves observed for an astronomical object
are usually caused by various processes at different timescales with
different coherences. The coherence coefficient $r(\Delta t)$
calculated by Eq.~(4) reflects not only the variation property
on the timescale $\Delta t$, but is also affected by that
on larger ones up to the total time period used in the calculation.
To limit the effect of larger timescales we divide the total observation
period into segments, each has a duration just 10$\Delta t$.
The resultant correlation coefficient should be then related to the
limited time scale region of $\Delta t - 10\Delta t$.

In our calculation for Cyg X-1 and GX 339-4 data,
for a given time bin $\Delta t$ the studied observation period is divided into
segments with a duration of 10$\Delta t$ each and the coherence coefficient
is calculated for each segment by using Eq.~(4).
The all segments are divided into ten groups with the segment numbers
in each group being nearly equal to each other.
We calculate 10 avarage coherence coefficients $r_k(\Delta t)$ for each group $k~(k=1,\dots,10)$
respectively and take their average $r(\Delta t)=\sum_{k=1}^{10}r_k(\Delta t)/10$
and the standard deviation $\sigma [r(\Delta t)]$ as the final result
of the observation. The Fourier coherence coefficients and their
average and the standard deviation of the average are also calculated from
the 10 time groups.
The results are shown in Figure 3.
For both Cyg X-1 and GX 339-4 in
the low state shown in the left and middle panels of Fig.~3, both
Fourier coherence and time domain coherence show similar trend of
change from near-perfect coherence at large time scale to low
coherence at small time scale. Two other black hole candidates, XTE
J1550-564 (Cui, Zhang and Chen 2000) and GRS 1915+105 (Ji et al.
2003), have Fourier coherence function with near-unity coherence at
large time scale and low coherence at small time scale. We also
apply the time domain technique to these sources and get similar
results. As the coherence is much more easily to be lost than
maintained, only when one linear physical process is involved could
coherence be preserved (Vaughan \& Nowak 1997). Any nonlinear
process or more than one independent linear processes involved could
degrade coherence. The near-perfect coherence at large time scale
may provide a strong evidence that there is a powerful physical
process in these sources with a large characteristic timescale
dominates the emission of photons in the studied energy bands.

It is a remarkable feature that all the values of time domain
coherence are very low in the whole studied timescale range for GX
339-4 in the high state, shown in the right panel of Fig.~3, indicating
 that there is only little relationship between variabilities of
2-5 keV and 5-7 keV emissions of GX339-4 in its high state.

\begin{figure}
\plotone{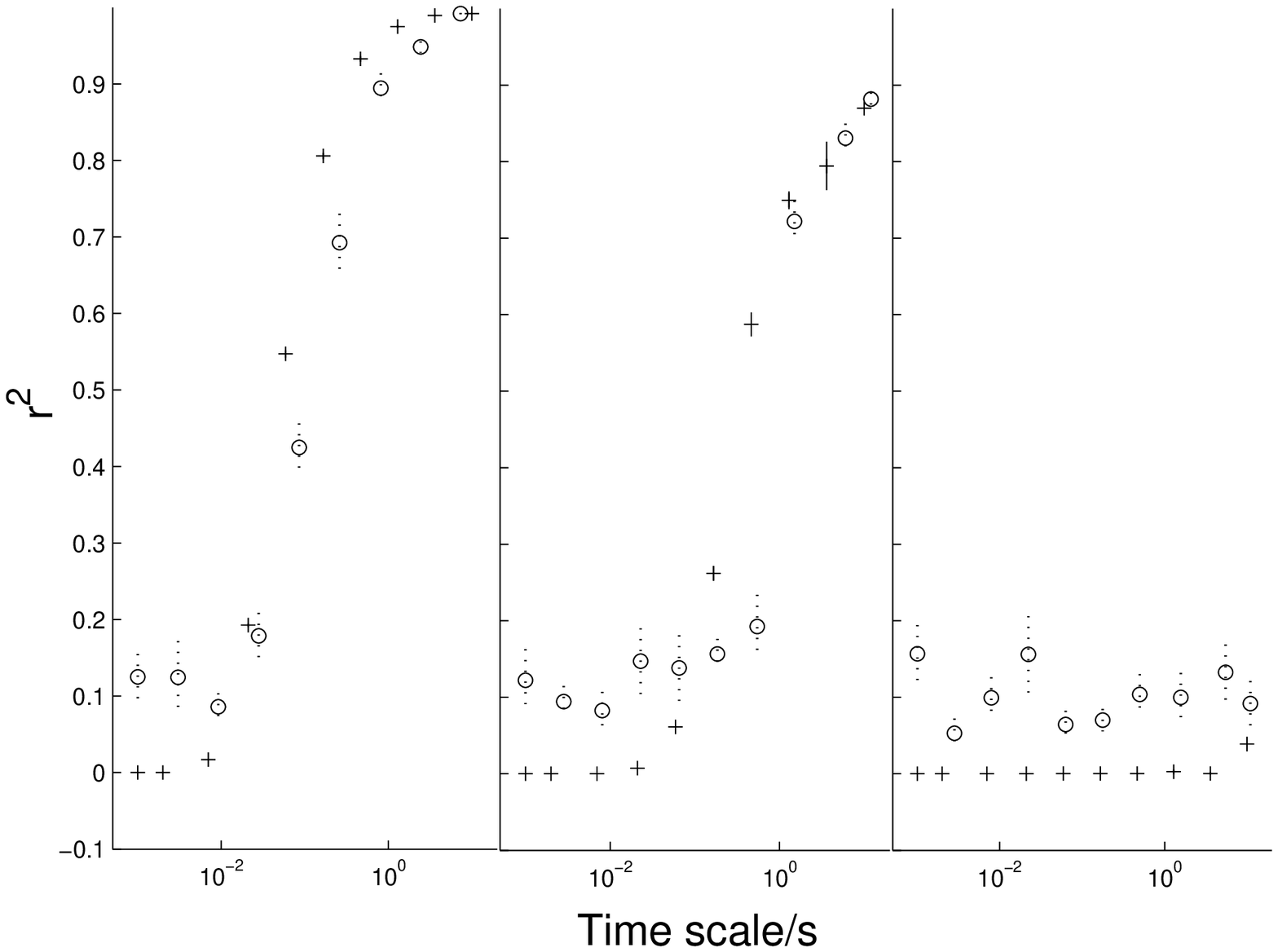} \caption{Variability coherence for black hole X-ray
binaries. {\sl Plus}: coherence obtained by the time domain
technique. {\sl Circle}: Fourier coherence function. {\sl Left
panel}: Cyg X-1 in the low state, PCA/$RXTE$ observation in
Dec-17-1996, 2-5 keV vs. 5-13 keV. {\sl Middle panel}: GX 339-4 in
the low state, PCA/$RXTE$ observation in Sep-19-1997, 2-6 keV vs.
6-8 keV. {\sl Right panel}: GX 339-4 in the high state, PCA/$RXTE$
observation in Jan-17-1998, 2-5 keV vs. 5-7 keV.
 \label{fig3}}
\end{figure}

To see whether the relationship between variabilities of two light curves
 coincides with the time domain coherence computed from them, we
simultaneously plot two observed light curves of Cyg X-1 in two energy bands
for different time scales at the same time period, shown in Figure~4.
From Fig.~4 one can see that the time domain coherence calculated by
Eq.~(4) can reflect the real strength of the linear relationship between
two light curves.

\begin{figure}
\plotone{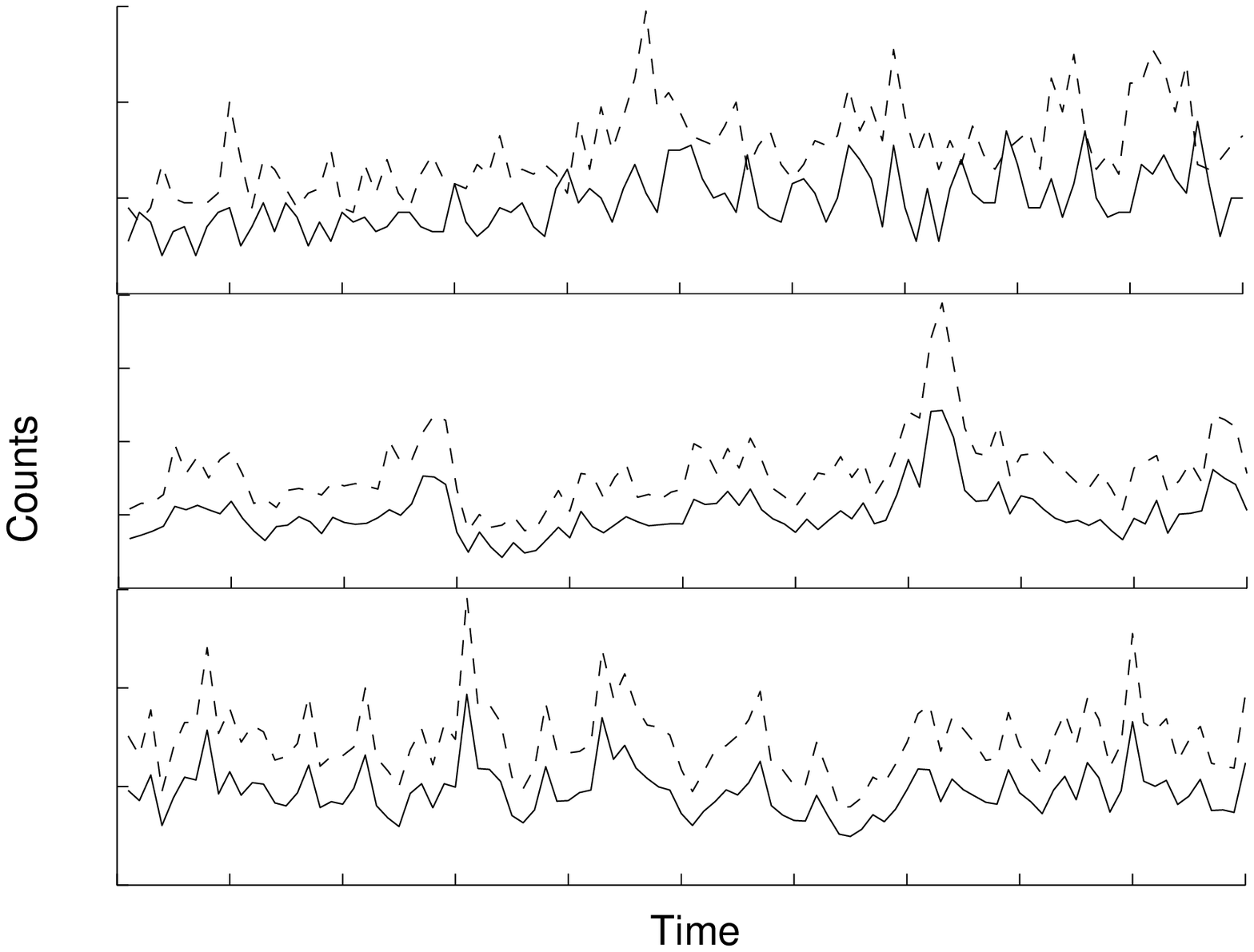} \caption{ Lightcurve segments of Cyg X-1 in the
low state, PCA/$RXTE$ observation in Dec-17-1996. $Solid~line$:
2-5 keV, $Dashed~line$: 5-13 keV. {\sl Top panel}: 10 ms time bin, time domain coherence
in Fig.~3 is $\sim 0.4$. {\sl Middle panel}: 100 ms time bin,
coherence $\sim 0.9$. {\sl Bottom panel}: 1000 ms time bin,
coherence is very near to unit.
 \label{fig4}}
\end{figure}

\section{Discussion}
The Fourier coherence function $r(f)$ of two lightcurves $x$ and $y$ is defined in the space of Fourier frequency
\begin{equation} r(f)=\frac{|\left<C(f)\right>|}{\sqrt{\left<|X(f)|^2\right>\left<|Y(f)|^2\right>}}~. \end{equation}
Where, $X(f)$ and $Y(f)$ are the Fourier transforms of $x$ and $y$ respectively,
the cross spectrum $C(f)=X^*(f)Y(f)$, and
angle brackets denote an average over the used segments of light curve.
If the ratio of two Fourier transforms at a frequency $f$,
$H(f)=X(f)/Y(f)$, is the same for all segments of the two processes,
or, equivalently
\begin{equation} y(t)=\int h(t-\tau)x(\tau)d\tau~, \end{equation}
then $r(f)=1$, the processes are said to be coherent at frequency $f$.
But the unit coherence in the time domain,
$r(\Delta t)=1$, indicates a linear correlation between the variabilities of intensity
in the two energy bands on the timescale $\Delta t$
\begin{equation}
y(i;\Delta t)=hx(i;\Delta t)~, \hspace{5mm} (i=1,\dots,N_T)~, \end{equation}
where $h$ is a constant during the observation period $T=N_T\Delta t$, which is a stronger constraint than
Eq.~(10) from an unit coherence, $r(f)=1$, in the frequency domain.
It is an advantage for Fourier coherence function that it can obtain
coherence at all frequencies simultaneously and exactly, because
amplitude at all frequencies could be obtained independently from
Fourier transformation of original lightcurve. Improved Fourier
coherence function (Vaughan \& Nowak 1997) can calculate coherence
of noisy lightcurves and uncertainty as well.
While the coherence coefficient calculated in the time domain by Eq.~(4) can be confined
in only a timescale region, $\Delta t - N\Delta t$ with $N$ being the number of time bins
in a calculated time segment, and the error of the average coherence coefficient is
estimated by the statistical deviation in coherence coefficients of
different time segments.

Spectral time lags are usually existed in observed processes.
To calculate the time domain coherence $r(\Delta t)$, one should firstly derive
their timescale spectrum of time lag, then make time lag correction for light curves
at the timescale $\Delta t$.
But, in practice, the coherence derived by the time domain
technique is not sensitive to the spectral time lag in analyzing
real observational data.
 From lightcurves of Cyg X-1 in its low
state in energy band 2-5 keV and 5-13 keV observed by PCA/$RXTE$, we
use the modified cross-correlate function (MCCF)  (Li, Feng \& Chen 1999; Li
et al. 2004) to calculate the spectral
time lag, the result is similar to
what obtained from Fourier analysis (Nowak et al 1999).
Then we calculate time domain coherences between the two lightcurves
with and without time lag correlation separately. The result shown in Fig.~5
does not indicate any essential effects of the spectral lag on the time
domain coherence. To see how seriously time lag could
affect results of time domain coherence, we make a modification
to the original observed lightcurves to produce an artificial deviation: the band
5-13 keV lags behind the band 2-5 keV by 10 times of time lag
measured from the observed lightcurves by MCCF, then calculate time domain
coherences between the modified lightcurves. As we can see in Fig.~5,
only a litter deviation can be found between the result from the original
lightcurves and the modified ones. We also test other data
used in this paper to see whether Fig.~5 is only an exception, the
results turns out to be that although the spectral time lag exists
in most data, it has just little influence to the coherence obtained
by the time domain technique.

\begin{figure}
\plotone{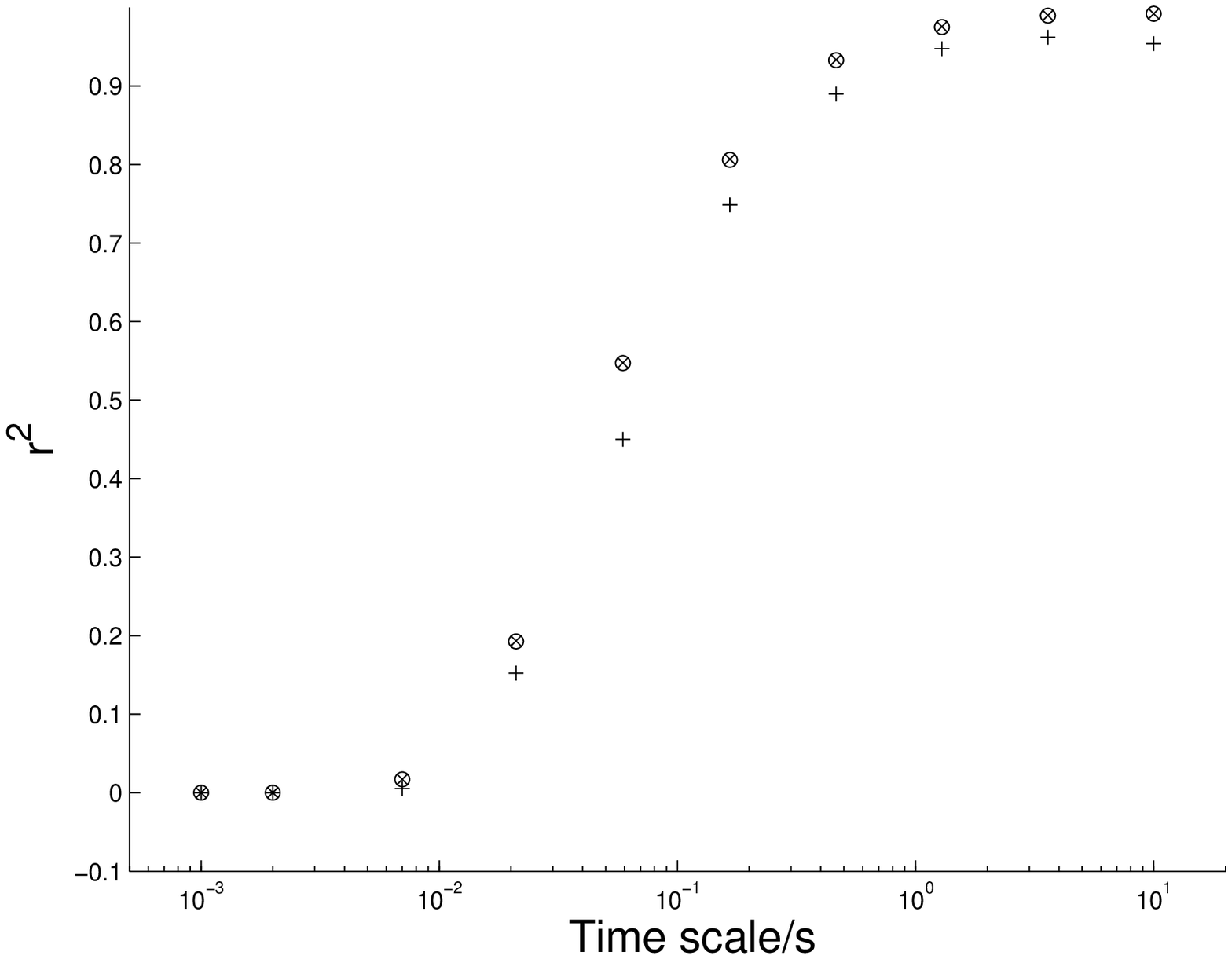} \caption{X-ray coherence obtained by the time
domain technique. Cyg X-1 in the low state, PCA/$RXTE$ observation
in Dec-17-1996, 2-5 keV vs. 5-13 keV. {\sl Circle}: from original
observed lightcurves. {\sl Cross}: from lightcurves after spectral
timelag correction. {\sl Plus}: from modified lightcurves whose timelag is
artificially amplified by 10 times. \label{fig5}}
\end{figure}

To understand reasons of insensitivity of time domain coherence to
spectral timelag in observed X-ray data, we make simulation studies.
Two 1000 s lightcurves with 1 ms time bin and a given correlation
coefficient $\rho=0.8$ (or expected coherence $r^2=0.64$) are
produced by sampling from the two-dimensional normal distribution
Eq.~(5) described in \S 2. One of the lightcurves is then
artificially shifted by 10 ms. We use MCCF to derive timescale
distribution of time lags and calculate time domain coherences
between the two lightcurves with and without timelag correction. The
results are shown in the left panel of Figure~6. We can see that,
though the effect of spectral timelag can not be neglected in
general, the difference of coherences obtained with and without
timelag correction is small at large timescale. For deriving time
domain coherences at large timescales, we need to reconstruct
original lightcurves with a larger time bin (see the details in \S
2), which impairs the effect of a timelag which is much smaller than
the time bin. Spectral timelags measured at a timescale are usually
much shorter than the timescale itself. For example, the X-ray time
lag of Cyg X-1 between 2-5 keV and 5-13 keV at the timescale of 1 s
(or frequency of 1 Hz) is only about 10 ms. From the left panel of
Fig.~6 we can see that the effect of timelag on time domain
coherence is small when timelag is much smaller than time bin.

\begin{figure}
\plotone{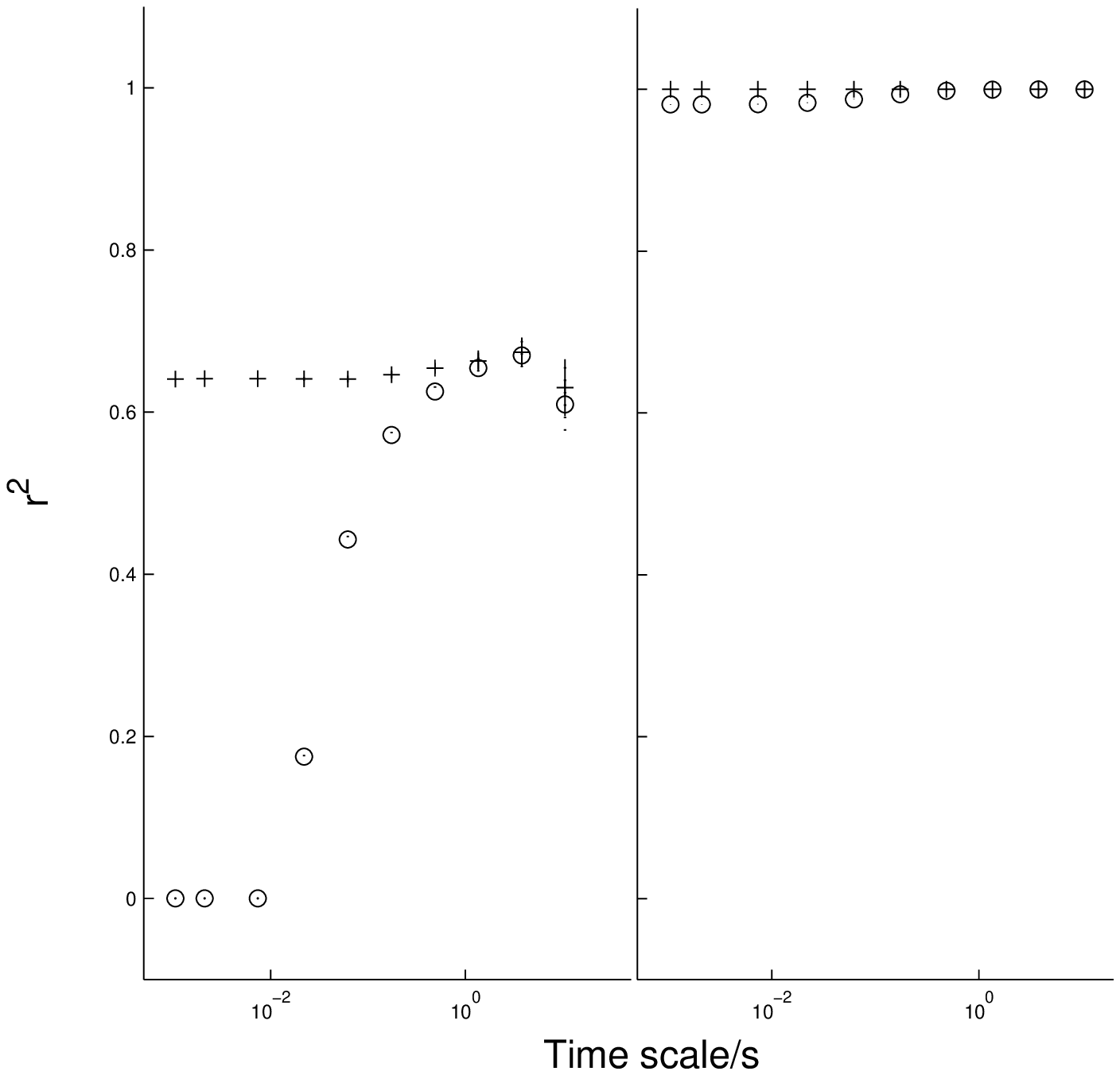} \caption{Spectral timelag effect on coherence
obtained by the time domain technique. {\sl Circle} : no timelag
correction. {\sl Plus}: after timelag correction. {\sl Left panel}:
coherence between two simulated lightcurves based on two-dimensional
normal distribution model with timelag 10ms at all time scale, the
given coherence is $r^2=0.64$. {\sl Right panel}: coherence between
two simulated lightcurves based on shot model with timelag 10ms at
all time scale.
 \label{fig6}}
\end{figure}

The correlation existed in
successive time bins in observed lightcurves may be another reason of the insensitivity
of time domain coherence to spectral timelag in observed X-ray data.
The right panel of Fig.~6 shows results from two simulated lightcurves
based on shot model with 10 ms timelag.
We produce a 1000 s lightcurve with 1 ms time bin consisted of
random shots similar to what observed from Cyg X-1 (Feng, Li \& Chen 1999).
The front of the shot is represented by a single exponential and
the time constant of both ascending and descending
front is 70 ms. The interval between two successive shots follows an
exponential distribution with an average of 450 ms.
The above produced lightcurve is shifted  by 10 ms to create the second one.
The right panel of Fig.~6 shows coherence spectra derived by
the two lightcurves with and without timelag correction.
From Fig.~6 we see that the timelag effect on time domain coherence
for lightcurves produced by shots (right panel)
is much smaller than that for lightcurves sampled from two-dimensional
normal distribution  (left panel).
This is because successive bins in a lightcurve consisted of shots have a strong
connection.

In summary, we propose a time domain technique in this paper to calculate
variability coherence for different timescales. In general, the time domain
coherence distribution is
consistent with the Fourier coherence function, but more reliable
for the case of low signal to noise ratio and more powerful to
reveal lower coherence.
 In practical data analysis, the
effect of spectral time lags on the time domain coherence can be
neglected. This technique provides an useful tool for temporal
analysis.

\acknowledgments
The referee is thanked for helpful comments and suggestions.
This research was supported by the National Natural
Science Foundation of China and  made use
of data obtained through the High Energy Astrophysics Science Archive Research
Center Online Service,
provided by the NASA/Goddard Space Flight Center.

\end{document}